\documentclass[%
reprint,
amsmath,amssymb,
aps,
floatfix,
]{revtex4-2}

\usepackage{graphicx}
\usepackage{dcolumn}
\usepackage{bm}
\usepackage[colorlinks=true,citecolor=blue,linkcolor=blue,urlcolor=blue]{hyperref}

\usepackage{tikz}
\usepackage{xcolor}
\usepackage{soul}

\newcommand{\orcidicon}[1]{\href{https://orcid.org/#1}{\includegraphics[height=\fontcharht\font`\B]{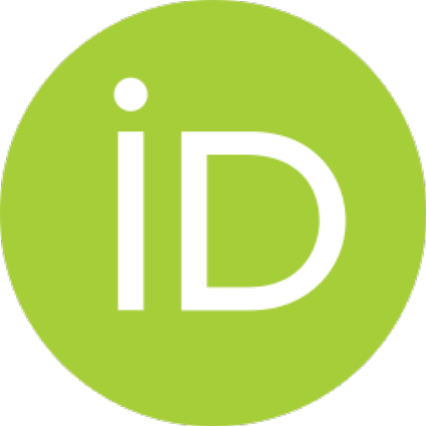}}}

\usepackage{mathrsfs}
\usepackage{pstricks}
\usepackage{color}
\usepackage{slashed}
\usepackage{epsfig}
\usepackage{amsfonts}
\def\beq{\begin{equation}}
\def\eeq{\end{equation}}
\def\bea{\begin{eqnarray}}
\def\eea{\end{eqnarray}}
\def\nn{\nonumber}

\def\Re{\textrm{Re}}

\def\x{\mathbf{x}}
\def\y{\mathbf{y}}

\def\q_perp{\mathbf{q}_{\perp}}
\begin{document}

\title{Restoration of a Spontaneously Broken Symmetry in a \\ 
Euclidean Quantum \texorpdfstring{$\lambda\varphi^{4}_{d+1}$} 
{} Model with Quenched Disorder}

\author{G. O. Heymans\,\orcidicon{0000-0002-1650-4903 }}
\email{Email address: olegario@cbpf.br}
\affiliation{Centro Brasileiro de Pesquisas F\'{\i}sicas, 22290-180 Rio de Janeiro, RJ, Brazil}
\author{N.~F.~Svaiter\,\orcidicon{0000-0001-8830-6925}}

\email{Email address: nfuxsvai@cbpf.br}
\affiliation{Centro Brasileiro de Pesquisas F\'{\i}sicas, 22290-180 Rio de Janeiro, RJ, Brazil}

\author{G.~Krein\,\orcidicon{0000-0003-1713-8578}}
\email{Email address: gastao.krein@unesp.br}
\affiliation{Instituto de F\'{i}sica Te\'orica, Universidade Estadual Paulista,\\
Rua Dr. Bento Teobaldo Ferraz, 271, Bloco II, 01140-070, S\~ao Paulo, SP, Brazil}
%


\begin{abstract}
We investigate the low-temperature behavior of a system in a spontaneously broken symmetry 
phase described by a Euclidean quantum  $\lambda\varphi^{4}_{d+1}$ model with quenched disorder. 
We study the effects of the disorder linearly coupled to the scalar field using a series 
representation for the averaged generating functional of connected correlation functions  
in terms of the moments of the partition function. To deal with the strongly correlated disorder 
in imaginary-time, we employ the equivalence between the model defined in a $d$-dimensional space 
with imaginary-time with the statistical field theory model defined on a space ${\mathbb R}^{d}\times S^{1}$ 
with anisotropic quenched disorder. Next, using  stochastic differential equations and fractional derivatives, 
we obtain  the Fourier transform of the correlation functions of the disordered system at tree level. 
In one-loop approximation, we prove that there is a denumerable collection of moments of the partition 
function that can develop critical behavior. Our main result is that, even with the bulk in the ordered phase,
there are many critical compactified lengths  that take each of the moments of the partition function from an 
ordered to a disordered phase. This is a sign of generic scale invariance emergence in the system. 
\end{abstract}


\maketitle



\section{Introduction}\label{intro}

The aim of this work is to discuss with quantum field theory techniques quenched disorder 
effects in systems at low temperatures in the spontaneously symmetry-broken phase. The classical 
and quantum descriptions of physical systems in the presence of quenched disorder are of fundamental 
importance. Static disorder, for instance, is present in many condensed-matter systems, such as disordered metals, 
impure semiconductors, and classical or quantum spin systems~\cite{malivro, livro1,livro3,livro4}. The effects 
of random couplings on second-order phase transitions in $d$-dimensional systems, driven by thermal and
disorder fluctuations, are controlled by the Harris criterion~\cite{harris}. Under coarse-graining fluctuations, 
which is a standard approach in the treatment of disordered systems, one can identify two distinguished 
behaviors of the system's criticality under disorder. Namely, if the correlation length exponent of 
the pure-system $\nu$ satisfies the inequality $\nu\geq\frac{2}{d}$, the effects of the disorder may be 
disregarded on the physics of large length scales.  Otherwise, if $\nu<\frac{2}{d}$ the disorder-induced 
fluctuations modify the critical behavior. In the latter case, the critical exponents must change under 
the coarse-graining procedure. 

A prototype model that can be studied as a continuous field in the presence of a random field is the binary 
fluid in a porous medium~\cite{broch}. When the binary-fluid correlation length is smaller than the porous radius, 
one has a system for studying finite-size effects in the presence of a surface field. When the binary fluid 
correlation length is much larger than that of the porous radius, the random porous can exert a random field 
effect. In the latter case, one can also introduce boundaries and thereby obtain a Casimir-like 
effect~\cite{ca1}, known as the statistical Casimir effect~\cite{cce16,bran,cce18}.

Recent experimental and theoretical advances have driven increased  interest in low-temperature physics and 
quantum phase transitions~\cite {zurek,zvyagin,heyl,gustavo,gustavo2}. Two questions dominate the interest
in the low-temperature physics of systems with quenched disorder~\cite{vojta1,belitz,vojta2,vojta3}:  
1) What is the effect of randomness in models at low temperatures in symmetry-broken phases? 2) What 
is the link between nonlocality (driven by anisotropic disorder) and the appearance of generic scale invariance 
in systems with continuous and discrete symmetry? We recall that generic scale invariance refers to
power-law decay of correlation functions in macroscopic regions of the system. It is well known that models 
with continuous symmetry can exhibit generic scale invariance due to the Goldstone theorem~\cite{gri}. 
Nevertheless, even in the case of discrete symmetry, the presence of quenched disorder also leads to generic 
scale invariance. Such behavior is in accord with Garrido~{\it et al.}~\cite{garrido}, who claim that a necessary, 
but not sufficient, condition for generic scale invariance is an anisotropic disorder. Later on, 
Vespignani and Zapperi~\cite{zapperi} showed that the breakdown of locality is essential to the emergence
of generic scale invariance. By now, it is a well-known fact that low temperatures in quenched 
disordered systems drive spatial nonlocality.  As such, one can rephrase the two questions with a single one: 
what is the link between low temperatures and generic scale invariance in systems with a quenched disorder?

In this work, we discuss the effects of a disorder field in a Euclidean quantum scalar $\lambda\varphi_{d+1}^{4}$ 
model at low temperatures in the broken symmetry phase. Criticality in this model is induced mainly by quantum and 
disorder fluctuations. The extreme situation, in which thermal fluctuations are absent, characterizes a quantum 
phase transition. In this case, the ground states of the system change in some fundamental way tuned by nonthermal 
control parameters~\cite{hertz,shakar,mvojta,sachdevbook}. For systems with quenched disorder, the physical quantity
of interest is the disorder-averaged free-energy~\cite{englert,lebo}. The random fields are described assuming 
some probability distribution on the space of realizations of the disorder. To calculate the disorder average,
different methods are used in the literature, such as the replica method~\cite{re,emery}, dynamic 
theories~\cite{dominicis, zip}, and supersymmetric approaches~\cite{efe1}. Another way to find the quenched 
free-energy is the distributional zeta-function method~\cite{distributional,distributional2,zarro1,zarro2,polymer,1haw,spin-glass}, 
which is the method we use in the present work to analyze the restoration of a spontaneously broken symmetry due 
to quantum and disorder effects. This field theory method allows us to obtain in a natural 
and unified way, locality breaking and generic scaling invariance due to an anisotropic quenched disorder.
In~the distributional zeta-function method, the disorder-averaged free-energy is represented by a series of moments 
of the partition function. In general, this leads to a complex free-energy landscape. However, a previous work~\cite{nami} 
pointed out that at criticality, a single moment dominates the series, a feature that allows us to concentrate 
the analysis on that single moment of the averaged free-energy. Here, we discuss how an anisotropy in the disorder changes 
the simpler scenario arising from an isotropic disorder.

To discuss quantum fluctuations effects in a system with the disorder in the broken symmetry phase at 
low temperatures, one can use the imaginary-time formalism~\cite{cb,jackiw, rep}. To study the low-temperature 
behavior of the model, since the disorder is strongly correlated in imaginary-time, we use the equivalence between 
a disordered Euclidean quantum $\lambda\varphi_{d+1}^{4}$ model with a classical model defined in a 
${\mathbb R}^{d}\times S^{1}$ space with anisotropic quenched disorder.
This model with a spatially nonuniform disorder has some similarities with the McCoy-Wu random Ising model~\cite{mc1,mc2}, 
an anisotropic two-dimensional classical Ising model with random exchange along one direction but uniform along the other;
see also Refs. \cite{d1,d2,d3}. Next, using  stochastic differential equations and fractional derivatives, 
we obtain the Fourier transform of the correlation functions of the disordered system at the tree level approximation. 
Going further, we implement a one-loop approximation to show that in the set of moments that define the quenched 
free-energy, there is a denumerable collection of moments that can develop critical behavior 
induced by quantum and disorder-induced fluctuations. With the bulk of the system in the ordered phase, a large 
number of critical compactified lengths take each of these moments from an ordered to a disordered phase. 
We demonstrate the link between the nonlocality and generic scale invariance for a system with quenched disorder.

We structure the paper as follows. In Sec.~\ref{sec:thermal mass} we revise an analytical regularization 
procedure that we use to regularize divergences appearing in the one-loop calculations. We exemplify the 
application of the regularization procedure with a pure  (no disorder)  scalar field model in the broken symmetry 
phase.  We review the distributional zeta-function method in Sec.~\ref{sec:frompathtoSPDE2}. We use a simple model, 
the random-mass scalar field model, to illustrate the essential steps to obtain the disorder-averaged
free-energy with the method. Section~\ref{sec:mod-qd} starts the original work of the paper. We obtain
the disorder-averaged free-energy for the model of interest, the Euclidean quantum 
$\lambda\varphi^{4}_{d+1}$ model with additive, anisotropic quenched disorder. The average free-energy is given as 
a series of the moments of the partition function, each of which is characterized by an effective action
of a multiplet of fields. The effective actions contain nonlocal terms generated by the anisotropic disorder. 
In Sec.~\ref{sec:frompathtoSPDE3},
we obtain dynamic correlation functions for the fields of a given moment. To obtain the correlation functions, 
we employ stochastic differential equations with additive white noise, with each equation driven by the effective 
action of the corresponding moment. To deal with the nonlocal terms that appear in these equations, we employ 
the method of fractional derivatives. In Sec.~\ref{sec:thermal mass2}, we compute in the one-loop approximation 
the effect of the disorder on the mass in the symmetry-broken phase. We also show numerical results for the 
mass for different values of the parameters of the model. We present our conclusions in Sec.~\ref{sec:conclusions}. 
The paper still contains one Appendix, in which we determine the set of 
critical moments. Throughout the paper we use $\hbar=c=k_{B}=1$ units.


\section{Analytic regularization of the one-loop approximation 
in a scalar field model}\label{sec:thermal mass}

The aim of this section is to review the analytic regulation method that we
use in Section~\ref{sec:thermal mass2} to discuss the local restoration of 
a low-temperature broken symmetry in a model with quenched disorder. 
To present the method, we discuss temperature effects in the low-temperature
broken symmetry phase of a real scalar field model without the 
disorder. The model contains self-interaction vertices $\rho_{0}\varphi^{3} 
+ \lambda_{0}\varphi^{4}$ and is defined on a ${\mathbb R}^{d}\times S^{1}$ 
space. We study the one-loop thermal corrections to the renormalized  
squared mass of the model. The contribution from the $\lambda_{0} \varphi^{4}$
vertex is a tadpole diagram, and the contribution of the $\rho_{0} \varphi^{3}$ 
vertex is a bubble with two vertices. We name the latter a self-energy contribution.

In the imaginary-time formalism, the action functional $S(\phi)$ is given by
the Euclidean action~\cite{e1,e2,e3}:
\begin{align}
S(\phi) &= \frac{1}{2}\!\int_{0}^{\beta} \!\! d\tau \!\! \int \!\! d^{d}x \, \left[ 
\phi(\tau,\x) \! \left( -\frac{\partial^{2}}{\partial\tau^{2}} 
- \Delta + \mu_{0}^{2}\right)\!\phi(\tau,\x) \right. \nonumber\\
& \left. \;\;\;  + \, \frac{\lambda}{2}\phi^{4}(\tau,\x) 
\right].
\label{99}
\end{align}
Here we omit a subscript $0$ indicating an unrenormalized field and coupling constant. 
The perturbative renormalization consists of the introduction of additive counterterms related 
to $\delta m^{2}$, $Z_{1}$, and $Z_{2}$. The partition function is given by the functional integral
\begin{equation}
Z=\int[d\phi]\,\, e^{-S(\phi)} ,
\end{equation}
\\ 
where $[d\phi]=\prod_{\tau,\x} d\phi(\tau,\x)$ is a functional measure. The field variables satisfy 
the periodicity condition $\phi(0,\x)=\phi(\beta,\x)$, where $\beta$ is the reciprocal temperature. 
All the finite temperature $n$-point Schwinger functions (or Euclidean correlation functions) can be expressed 
as moments of this measure:
\begin{align}\label{eq:correlationpure}
\hspace{-0.2cm}\langle\phi(\tau_{1},\x_{1}) \cdots \phi(\tau_{n},\x_{n})\rangle \! = \! 
\frac{1}{Z}\!\int\! [d\phi] \prod_{i=1}^{n}\!\phi(\tau_{i},\x_{i}) 
\, e^{- S(\phi) }.
\end{align}

As usual, to generate the correlation functions of the model by functional derivatives 
one can introduce an external source, defining the generating functional of correlation 
functions $Z(j)$. From this functional, one can define the generating functional of connected 
correlation functions $W(j)$. Performing a Legendre transform one gets the generating functional 
$\Gamma(\varphi)$  of vertex functions. The renormalization conditions are imposed over the 
vertex functions. Starting from a new model with a negative sign of the mass squared 
$(\mu_{0}^{2}\rightarrow -\mu_{0}^{2})$ it is possible to show that quantum and thermal effects 
restore the symmetry-broken at the tree level $(\phi\rightarrow -\phi)$, where the ground state 
is unique. Above the critical temperature, one can write the thermal correction to the mass 
squared as
\begin{equation}
m^{2}_{R}(\beta)=-\mu_{0}^{2}+\delta \mu^{2}+\Delta m^{2}(\beta),
\end{equation}
where $\delta \mu^{2}$ is the usual mass counterterm that must be introduced in the 
renormalization procedure to obtain a finite result. The situation of temperatures below 
the critical temperature will be discussed in the following. The result of symmetry restoration 
is easily obtained at the one-loop level. We note that beyond perturbation theory, one can 
use the Dyson-Schwinger equation to obtain a nonperturbative result~\cite{gino}. To study 
spontaneous symmetry breaking beyond the tree level, it is well known that for a translationally 
uniform system, one can use the effective potential, the free-energy per unit volume~\cite{cw}. However,
since our aim is to investigate a model with a spatially nonuniform disorder, another 
method must be used.

Let us suppose that the system is in the ordered phase, at low temperatures, with 
some nonzero vacuum expectation value $v=\sqrt{\mu_{0}^{2}/{\lambda}}$, using the quantum 
field theory terminology. Shifting the field as $\varphi=\phi-v$, the new theory has an 
effective mass squared $m_{0}^{2}=3\lambda v^{2}-\mu^{2}_{0}$ and self-interaction vertices 
$\rho_{0}\varphi^{3}$ and $\lambda_{0}\varphi^{4}$, where $\rho_{0}=2\lambda v$ and 
$\lambda=\lambda_{0}$. In the case of a continuous second-order phase transition, one 
has to find the corrections to the mass. The one-loop correction for the spontaneous 
symmetry-broken phase can be easily discussed. Tadpole and bubble diagrams 
give the first non trivial contributions, we denote them  $\Delta m^{2}_{1}(\beta)$ and 
$\Delta m^{2}_{2}(\beta)$: 
\begin{align}
m_{R}^{2}(\beta) = m_{0}^{2} + \delta m^{2}_{0} 
+ 3\, \Delta m^{2}_{1}(\beta) 
+ 9\,\Delta m^{2}_{2}(\beta) ,
\end{align}
where $3$ and $9$ are symmetry factors, and $\delta m^{2}_{0}$ is a $d$-dependent mass 
counterterm. We omitted a $d$ dependence in the $\Delta m(\beta)$ functions to lighten 
the notation. Let us first discuss the tadpole contribution. 

Different procedures are used in the literature to evaluate the Matsubara sum of the tadpole. 
One can use a method where the Matsubara frequency sum separates into temperature-independent 
and temperature-dependent parts.  An alternative procedure is to use a mix between the dimensional 
and analytic regularization procedure \cite{dim1,dim2,thooft,physica}. Here we will use 
an analytic regularization procedures, where the number of dimensions of the space is not a complex
continuous variable  \cite{bbb}. A detailed study comparing an analytic regularization procedure 
and a cutoff method in the Casimir effect can be found in Refs. \cite{ca12,ca13,ca14,ca15}. 
The analytic regularization procedure aims to replace divergent integrals with analytic functions 
of certain regularization parameters.

We denote the thermal contribution from the tadpole, after analytic continuation, by 
$\Delta m^{2}_{1}(\beta,\mu,s)|_{s=1}$. Performing the angular part of the integral over the continuous 
momenta of the noncompact $d$-dimensional space, for $s\in \mathbb{C}$,  the $\Delta m^{2}_{1}(\beta)$ 
quantity can be written as $\Delta m^{2}_{1}(\beta,\mu,s)$, where 
\begin{align}
\Delta m^{2}_{1}(\beta,\mu,s) &= \frac{\lambda(\mu,s)\,\beta}{2^{d+1}\pi^{\frac{d}{2}+1}\Gamma\bigl(\frac{d}{2}\bigr)} 
\nonumber \\
&\hspace{-1.0cm}\times \int_{0}^{\infty}\! dp \, p^{d-1}
\sum_{n\in \mathbb{Z}}\left[\pi n^{2}
+ \frac{\beta^{2}}{4\pi}\Bigl(p^{2}
+ m_{0}^{2}\Bigr) \right]^{-s} ,
\label{delta_m1}
\end{align}
with $\lambda(\mu,s)=\lambda_{0}(\mu^{2})^{s-1}$, where $\mu$ has mass dimension. The function 
$\Delta m^{2}_{1}(\beta,\mu,s) $ is defined in the region where the above integral converges, 
$\Re(s)>s_{0}$. 

The self-energy contribution to the mass, $\Delta m^{2}_{2}(\beta)$,  can be obtained 
from the tadpole as: 
\begin{equation}
\Delta m^{2}_{2}(\beta)=\biggl[-\frac{\rho^{2}(\mu,s)}{\lambda(\mu,s)} \, 
\Delta m^{2}_{1}(\beta,\mu,s) \biggr]_{s=2}, 
\end{equation}
where $\rho(\mu,s)=\rho_{0}(\mu^{2})^{s-2}$. Therefore, one can concentrate on the 
$\Delta m^{2}_{1}(\beta,\mu,s) $ function. 

After a Mellin transform and reordering of some quantities, we can write 
$\Delta m^{2}_{1}(\beta,\mu,s)$ as:
\begin{align}
&\Delta m^{2}_{1}(\beta,\mu,s) = 
\frac{\lambda(\mu,s)}{2\pi\Gamma(\frac{d}{2})\Gamma(s)}\biggl(\frac{1}{\beta}\biggr)^{d-1}
\nonumber\\
&\hspace{0.15cm}
\times \int_{0}^{\infty}\!\! dr\,r^{d-1} \!\!\int_{0}^{\infty}\!\!dt\,t^{s-1}
\sum_{n\in \mathbb{Z}} e^{-\left(\pi\,n^{2}+r^{2}+ m_{0}^{2}\beta^{2}/{4\pi}\right)t} ,
\label{m1-mu-s}
\end{align}
where we made the change of variable $r^{2}={\beta^{2}p^{2}}/{4\pi}$. 
The integral over $r$ is straightforward. We represent the sum over $n\in \mathbb{Z}$ 
by $\Theta(v)$: 
\begin{equation}
\Theta(v)=\sum_{n\in \mathbb{Z}} e^{-\pi\,n^{2}v},
\end{equation}
which is an example of a modular form. Then, we split the $t$ integral into two:
\begin{align} 
\hspace{-0.2cm}\Delta m^{2}_{1}(\beta,\mu,s) &\!=\! C_{d}(\beta,\mu,s) \Biggl[
\int_{0}^{1}\!\!dt\,t^{s-\frac{d}{2}-1} 
\, e^{- m_{0}^{2}\beta^{2}t/{4\pi}}  \, \Theta(t) \nn \\
& + \,  
\int_{1}^{\infty}\!\!dt\,t^{s-\frac{d}{2}-1}
e^{ - m_{0}^{2}\beta^{2}t/{4\pi} }\,\Theta(t)\Biggr], 
\end{align}
with $C_{d}(\beta,\mu,s)$ defined as
\begin{equation}
C_{d}(\beta,\mu,s)=\frac{\lambda(\mu,s)}{4\pi\Gamma(s)}\biggl(\frac{1}{\beta}\biggr)^{d-1}.
\end{equation}
Next, by making a change of variable $t \rightarrow 1/t$ in the first integral
and using the modular property of the $\Theta$ function, 
\begin{equation} 
\Theta(v) = \frac{1}{\sqrt{v}}\Theta\left(\frac{1}{v}\right), 
\end{equation}
one can write $\Delta m^{2}_{1}(\beta,\mu,s)$ as a sum of four integrals:
\begin{align}
\Delta m^{2}_{1}(\beta,\mu,s) &= C_{d}(\beta,\mu,s) 
\Bigl[2I_{d}^{(1)}(\beta,s)  + 2I_{d}^{(2)}(\beta,s) \nn \\ 
& + \, I_{d}^{(3)}(\beta,s)+I_{d}^{(4)}(\beta,s)\Bigr],
\end{align}
where
\begin{align} 
I_{d}^{(1)}(\beta,s) &= \int_{1}^{\infty} dt\,t^{s-\frac{d}{2}-1} \, 
\, e^{ - m_{0}^{2}\beta^{2}t/{4\pi} } \; \psi(t), \\ 
I_{d}^{(2)}(\beta,s) &= \int_{1}^{\infty}dt\,t^{-s+\frac{d}{2}-\frac{1}{2}} \, 
e^{- m_{0}^{2}\beta^{2}/{4\pi t} } \; \psi(t), \\
I_{d}^{(3)}(\beta,s) &= \int_{1}^{\infty}dt\,t^{s-\frac{d}{2}-1}
e^{ - {m_{0}^{2}\beta^{2}t}/{4\pi} } , \\
I_{d}^{(4)}(\beta,s) &= \int_{1}^{\infty}dt\,t^{-s+\frac{d}{2}-\frac{1}{2}}
e^{ - {m_{0}^{2}\beta^{2}}/{4\pi t} } , 
\end{align}
in which $\psi(v)$ is given by
\begin{equation} 
\psi(v) = \sum_{n=1}^{\infty} e^{-\pi n^{2}v} = \frac{1}{2}\left[ \Theta(v)-1 \right]. 
\end{equation}

Now we can use the standard result that a function that is analytic on a domain 
$\Omega\subset \mathbb{C}$ has a unique extension to a function defined in 
$\mathbb{C}$, except for a discrete set of points. Using the fact that 
$\psi(t)=O(e^{-\pi t})$ as $t\rightarrow \infty$, the integrals $I_{d}^{(1)}(s,\beta)$ 
and $I_{d}^{(2)}(s,\beta)$ represent everywhere regular functions of $s$ 
for $m_{0}^{2}\beta^{2}\in \mathbb{R}_{+}$. The upper bound ensures uniform convergence 
of the integrals on every bounded domain in $ \mathbb{C}$. On the other hand, at low
low temperatures, the integrals $I_{d}^{(3)}(s,\beta)$ and $I_{d}^{(4)}(s,\beta)$ 
are finite too. Therefore, one can take the limit $s \rightarrow 1$ to obtain the
tadpole contribution to the thermal correction to the mass.  
Note that the thermal correction from the self-energy contribution is also finite; 
recall that to obtain this contribution we have to evaluate the four
integrals for $s=2$. We stress that these results are valid only in the
low-temperature situation. 

We note that we are left with an ultraviolet divergence that needs to be normalized.
The divergence comes from the integral $I_{d}^{(4)}(\beta,s)$. The renormalization is 
done by introducing a mass counterterm of the form  $-\delta m_{0}^{2} = C_{d}(\beta,\mu,s) 
I_{d}^{(4)}$.  This is a temperature-dependent counterterm coming from the subtraction 
at zero momentum of the self-energy diagram. Going beyond one-loop approximation, one 
can show that the counterterms of a finite temperature field theory are the same as the 
zero temperature theory. The final result is then: 
\begin{align}
\Delta m^{2}_{1}(\beta) &= C_{d}(\beta,1) 
\biggl[2I_{d}^{(1)}(\beta,1)+2I_{d}^{(2)}(\beta,1) \nn \\
& + \, I_{d}^{(3)}(\beta,1) + I_{d}^{(4)}(\beta,1)\biggr],
\\
\Delta m^{2}_{2}(\beta) & = -  C_{d}(\beta,2)  \, \frac{\rho_{0}^{2}}{\mu^2\lambda_{0}} 
\biggl[2I_{d}^{(1)}(\beta,2) + 2I_{d}^{(2)}(\beta,2) \nn \\ 
& + \, I_{d}^{(3)}(\beta,2) + I_{d}^{(4)}(\beta,2)\biggr],
\end{align}

\begin{figure}[htb]
\includegraphics[scale=0.67]{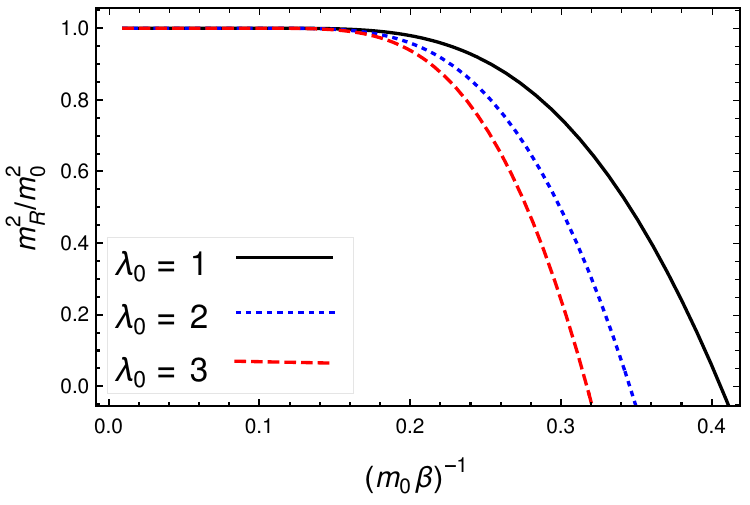}
\label{fig:1}
\caption{The squared mass as a function of $(m_0\beta)^{-1}$, for different values of the coupling 
constant $\lambda_0$ for $d=3$. We set $\mu^2=m_0^2$.}
\end{figure}

Finally, the critical temperature of this pure-system is given by the value
of $\beta$ for which the mass squared $m^2_R(\beta)$ vanishes. Figure~1 
presents the numerical results for for $d=3$ as a function of $m_0$ and 
selected values of $\lambda_0$.

\section{Distributional zeta-function method}
\label{sec:frompathtoSPDE2}

The aim of this section is to review the distributional zeta-function method~\cite{distributional,distributional2}
used to obtain the disorder-averaged free-energy. We use a simple Ginzburg-Landau type of model 
to explain how the distributional zeta-function method works. Specifically, we employ the random-mass
model with a $\lambda \phi^4$ term and show how one can obtain the free-energy landscape of the model.
The simplicity of the random-mass model stems from the fact that the disorder average modifies only 
this non-Gaussian term in the effective action, as explained in Ref.~\cite{zarro2}. 

The partition function of the model for one disorder realization in the presence of an external 
source $j(\textbf{x})$ is given by: 
\begin{equation}
\hspace{-0.2cm}Z(j,h)\!=\!\int\![d\phi]\, \exp\!\left[ -S(\phi,h) 
\!+\! \int \! d^{d+1}x\, j(\textbf{x})\phi(\textbf{x})\right]\!,
\label{eq:disorderedgeneratingfunctional}
\end{equation}
where $[d\phi] =\prod_{\textbf{x}} d\phi(\textbf{x})$ is a functional measure, and the action functional 
in the presence of the disorder is  
\begin{equation}
S(\phi,h)=S(\phi)+
\int d^{d+1}x \,
h(\textbf{x})\phi^{2}(\textbf{x}).
\label{eq:spe1}
\end{equation}
Here, $S(\phi)$ is the pure-system action of the Ginzburg-Landau type, and  $h(\textbf{x})$ is 
a quenched random field. 

In a general situation, one can model a disordered medium by a real random field $h(\x)$ in 
$\mathbb{R}^{d}$ with $\mathbb{E}[h(\x)]=0$ and covariance $\mathbb{E}[h(\x)h(\y)]$, where 
$\mathbb{E}[\cdots]$ specifies the mean over the ensemble of realizations of the disorder. 
As in the pure-system case, one can define the  generating functional of connected correlation 
functions for one disorder realization, $W(j,h)=\ln Z(j,h)$, or the system's free-energy for one 
disorder realization. From $W(j,h)$, one can obtain the quenched free-energy by performing the 
average over the ensemble of all disorder realizations. 

The physical quantity of interest is the disorder-averaged free-energy functional $\mathbb{E}[W(j,h)]$:
\begin{equation}
\mathbb{E}\bigl[W(j,h)\bigr]=\int\,[dh]P(h)\ln Z(j,h),
\label{eq:disorderedfreeenergy}
\end{equation} 
where $[dh] = \prod_{\x} dh(\x)$ is a functional measure and $[dh]P(h)$ is the probability distribution 
of the disorder field. For a general disorder probability distribution, the distributional zeta-function, 
$\Phi(s)$, is defined as~\cite{distributional,distributional2}:
\begin{equation}
\Phi(s)=\int [dh]P(h)\frac{1}{Z(j,h)^{s}}.
\label{pro1}
\vspace{.2cm}
\end{equation}
For $s\in \mathbb{C}$, this function is defined in the region where the above integral 
converges. One defines the complex exponential $n^{-s}=\exp(-s \log n)$ for $\log n\in\mathbb{R}$.
As proved in Refs.~\cite{distributional,distributional2}, $\Phi(s)$ is defined for $\Re(s) 
\geq 0$. Therefore, the integral is defined in the half-complex plane, and an analytic continuation 
is unnecessary.We have that 
\begin{equation}
\mathbb{E}\bigl[W(j,h)\bigr] = - \frac{d\Phi(s)}{ds}\Bigg|_{s=0^{+}}, \,\,\,\,\,\,\,\,\,\, \Re(s) \geq 0.  
\end{equation}

Using the Euler's integral representation for the gamma function, we get

\begin{equation}
    \Phi(s) = \frac{1}{\Gamma(s)}\int[dh]P(h)\int_{0}^{\infty}dt\,t^{s-1} e^{-Z(j,h)t}.
\end{equation}
The series expansion of the exponential in the previous integral has a uniform convergence for each $h$ in the domain $[0,a]$. To perform the integration term by term we need to split the integral into the contribution that is uniformly convergent, $t \in [0,a]$, and into the one that is not. The first contribution is a sum over all the integer moments of the partition function, $\mathbb{E}[Z^k(j,h)]$, while the second one is a constant which can be taken small as desired.

Therefore, one can show that the average free-energy can be represented by the following
series of the moments of the partition function~\cite{distributional}:
\begin{align}
\mathbb{E}\bigl[W(j,h)\bigr]&=\sum_{k=1}^{\infty} \frac{(-1)^{k+1}a^{k}}{k k!}\,\mathbb{E}\,[(Z(j,h))^{\,k}]\nonumber  \\
& - \, \ln(a)-\gamma+R(a,j),
\label{m23e}
\end{align}
\noindent where $a$ is a dimensionless arbitrary constant, $\gamma$ is the Euler-Mascheroni 
constant~\cite{abramowitz}, and $|R(a)|$ given by
\begin{equation}
R(a,j)=-\int [dh]P(h)\int_{a}^{\infty}\,\dfrac{dt}{t}\, e^{ -Z(j,h)t} .
\end{equation}
For large $a$, $|R(a)|$ is small, therefore, the dominant contribution to the average generating 
functional of connected correlation functions is given by the moments of the partition function 
of the model. We absorb $a$ in the functional measure and assume it is large.

We assume a Gaussian form for the probability distribution of the disorder field $[dh]\,P(h)$: 
\begin{equation}
P(h) = p_{0}\, \exp\left[ - \frac{1}{2\varrho^{2} } \int \! d^{d+1}x \; (h(\x))^{2}\right],
\label{dis2}
\end{equation}
where $\varrho$ is a positive parameter and $p_{0}$ is a normalization constant. In this case, we 
have a delta correlated disorder:
\begin{equation}
\mathbb{E}[{h(z,\x) h(z',\y)}] = \varrho^{2}\,\delta^{d}(\x-\y).
\label{delta-correl}
\end{equation}
After integrating the disorder, one obtains that each moment of the partition function  
$\mathbb{E}\,[Z^{\,k}(j,h)]$  can be written~as:
\begin{equation}\label{eq:29}
\mathbb{E}\,[Z^{\,k}(j,h)]=\int\,\prod_{i=1}^{k} \, [d\phi_{i}^{(k)}]\, e^{-S_{\textrm{eff}}(\phi_{i}^{(k)},j_{i}^{(k)})},
\end{equation}
where $S_{\textrm{eff}}(\phi_{i}^{(k)},j_{i}^{(k)})$ is obtained via the coarse-graining procedure, a standard procedure
in the literature~\cite{livro3,livro4}. In the above equation the superscript $(k)$, in $\phi_i^{(k)}$, identifies the term of the series expansion given by Eq. (\ref{m23e}) and the subscript $i$ is the component of the $k^{\mathrm{th}}$ multiplet. The $\prod_{i=1}^{k}[d\phi_{i}^{(k)}]$ represents a product of functional measures.

Specifically, for a Ginzburg-Landau model with $\lambda \phi^4$ and $\rho \phi^6$ terms, 
after performing the noise average, one obtains the effective action~\cite{zarro2}:
\begin{align}\label{eq:effectivehamiltonian}
&\hspace{-0.25cm}S_{\text{eff}}(\phi_{i}^{(k)}) \!=\!\! \int\!\! d^{d+1}x\, \sum_{i=1}^{k}  \left[\frac{1}{2} \phi_{i}^{(k)}(\x)
\Bigl(-\Delta\,+m_{0}^{2}\Bigr)\phi_{i}^{(k)}(\x)\right. \nonumber \\
&\hspace{-0.25cm} + \left.\frac{1}{4}\!\sum_{j=1}^{k} g_{ij} \left(\phi_{i}^{(k)}(\x)\right)^2\!\!\left(\phi_{j}^{(k)}(\x)\right)^{2}
\!+\frac{\rho}{6} \left(\phi_{i}^{(k)}(\x)\right)^6\right]\!,
\end{align}
where the coupling constants $g_{ij}$ are given by $g_{ij} = (\lambda \delta_{ij}-\varrho^2)$. The $\phi^6$ term 
is necessary to stabilize a ground state for the system since the disorder average introduces a $\phi^4$ term
with negative coupling for large moments of the partition function. If, for simplicity, we use $\phi_{i}^{(k)}(\x)=\phi_{j}^{(k)}(\x)$, the effective action 
takes the form: 
\begin{align}\label{eq:effectivehamiltonian2}
S_{\textrm{eff}}\left(\phi_{i}^{(k)}\right)&=\int d^{d+1}x \sum_{i=1}^{k}\left[\frac{1}{2}\phi_{i}^{(k)}(\x)
\Bigl(-\Delta+m_{0}^{2}\Bigr)\phi_{i}^{(k)}(\x)\right. \nonumber \\
&\hspace{-0.75cm} +\left.\frac{1}{4}\bigl(\lambda-k\varrho^2\bigr)\left(\phi_{i}^{(k)}(\x)\right)^{4}+\frac{\rho}{6}
\left(\phi_{i}^{(k)}(\x)\right)^{6}\right].
\end{align}
At this moment one comment is in order. Using such a choice for the functional space the calculations are simplified. Also we recover the expected behavior of the free-energy landscape discussed in the literature. In other words, the series over $k$ describes a multi-valley structure, which is a typical feature 
of glassylike phases in complex systems.

The next section starts with the main topic of this work, the low-temperature behavior of a scalar field model 
with quenched disorder. Our main interest is situations where the thermal fluctuations are negligible and 
disorder-induced and quantum fluctuations dominate.

\section{Euclidean Quantum \texorpdfstring{$\lambda\varphi^{4}_{d+1}$} {} model with Quenched Disorder}
\label{sec:mod-qd}

We consider a scalar field model with a $\lambda\varphi^{4}_{d+1}$ interaction defined in a 
${\mathbb R}^{d}\times S^{1}$ space. The aim is to discuss the model's broken symmetry phase at low 
temperatures in the presence of disorder linearly coupled disorder. In the imaginary-time formalism,
the action of the model is given by: 
\begin{equation} 
S(\phi,h) = S(\phi)+\int_{0}^{\beta} d\tau\int d^{d}x\,h(\x)\phi(\tau,\x). 
\end{equation}
where $S(\phi)$ is the pure-system action, given in Eq.~(\ref{99}). 

Using the distributional zeta-function formalism, reviewed in the previous section, one can find 
the imaginary-time effective action for each moment of the partition function in the presence of 
the external source. The new field variables appearing in those moments are also assumed to satisfy 
the periodicity condition in imaginary-time, \textit{i.e.}, $\phi_{i}^{(k)}(0,\x)=\phi_{i}^{(k)}(\beta,\x)$. 
%
%
Defining again $\varphi_i = \phi_i - v$, one obtains the traditional spontaneous symmetry-breaking  scenario 
discussed in the previous section.

In Euclidean scalar quantum field theories,  finite temperature effects and periodic boundary conditions 
in one of the spatial dimensions are on the same footing. That is, the scalar theory defined on a ${\mathbb R}^{d}\times S^{1}$
space is formally equivalent to the thermal scalar field theory since the momentum variable associated with one
of the spatial coordinates runs over discrete values multiples of ${2\pi}/{L}$, where $L$ is the length of one of 
the compactified spatial dimensions, which is similar to the Matsubara frequencies when one replaces $L$ with $\beta$. 

The behavior of a system in which quantum and disorder fluctuations dominate can be described either 
by a $d$-dimensional Euclidean quantum field theory (with $\beta\rightarrow\infty$) or a statistical 
field theory in $(d+1)$ dimensions. In this work, we use this equivalence to avoid potential 
confusion with the extra time variable in the stochastic differential equations used to describe temporal evolution 
of the system. In addition, we assume that the disorder field is strongly correlated in the compactif
ied dimension (imaginary-time). This assumption implies in a spatially nonuniform disorder field in the $(d+1)$-dimensional
classical Euclidean field theory, which we assume delta-correlated:
\begin{equation}
\mathbb{E}[{h(z,\x)h(z',\y)}] = \varrho^{2}\delta^{d}(\x-\y).
\end{equation}
In this case, we get a $(d+1)$ Euclidean space with fields obeying periodic boundary conditions
in one spatial coordinate. As in the finite temperature case, the series representation of the quenched free-energy
leads to one effective action for each moment of the partition function, namely: 
\begin{widetext}
\bea
S_{\textrm{eff}}\left(\varphi_{i}^{(k)},j_{i}^{(k)}\right)=&&\frac{1}{2}\int_{0}^{L}dz\int d^{\,d}x
\Biggl[\sum_{i=1}^{k}\biggl(\varphi_{i}^{(k)}(z,\x)\bigl(-\frac{\partial^{2}}{\partial z^{2}}
-\Delta+m_{0}^{2}\bigr)\varphi_{i}^{(k)}(z,\x) + \rho_{0}\bigl(\varphi_{i}^{(k)}(z,\x)\bigr)^{3}
+ \frac{\lambda_{0}}{2}\bigl(\varphi_{i}^{(k)}(z,\x)\bigr)^{4}\biggr)\Biggr]\nonumber \\
&&-\frac{1}{2}\int_{0}^{L}dz\int d^{d}x\sum_{r,s=1}^{k}\varphi_{r}^{(k)}(z,\x)j_{s}^{k}(z,\x)
-\frac{\varrho^{2}}{2 L^2}\int_{0}^{L} dz\int_{0}^{L} dz'\int d^{d}x\sum_{r,s=1}^{k}
\varphi_{r}^{(k)}(z,\x) \varphi_{s}^{(k)}(z',\x),\nonumber \\
\label{SeffA}
\eea
\end{widetext}
with $\varphi_{i}^{(k)}(0,\x)=\varphi_{i}^{(k)}(L,\x)$ and $j_{i}^{(k)}(0,\x)=j_{i}^{(k)}(L,\x)$. 
One sees that the last term in this expression is spatially nonlocal. Such a nonlocal contribution also appears 
in other models. For example, using renormalization group techniques and the replica trick in a random-mass model, 
Refs.~\cite{gr,gr2} find nonisotropic scaling behavior. In our approach, because the disorder is anisotropic, we 
find similarly that the critical behavior of the system is different for the compactified and noncompactified
directions.

In order to avoid unnecessary complications, and for practical purposes, we assume the following configuration of the scalar fields $\phi^{(k)}_{i}(\tau,\x)=\phi^{(k)}_{j}(\tau,\textbf{x})$
in the function space and also $j_{i}^{(k)}(\tau,\textbf{x})=j_{l}^{(k)}(\tau,\textbf{x})$ $\forall \,i,\,j$. For simplicity we redefined  $\phi'^{\,(k)}_{i}(\tau,\x)= 
 \frac{1}{\sqrt{k}}\phi^{(k)}_{i}(\tau,\x)$ and $\lambda'_{0}=\lambda_{0}k$. All the terms of the series have the same structure and one minimizes each term of the series one by one.  

In the next section, we use the fact that the pure model can be formally expanded in a perturbative series 
starting from the Gaussian model. We evaluate the temporal correlation of the disordered model in the Gaussian 
approximation ($\rho_{0}=0$ and $\lambda_{0}=0$). 


\section{Linear nonlocal Stochastic Differential Equations and Fractional derivatives} \label{sec:frompathtoSPDE3}

Here we discuss the consequences of introducing randomness in a quantum system at low temperatures 
in the spontaneously broken phase in the model discussed above, a statistical field theory with anisotropic 
disorder. Instead of computing correlation functions directly from the functional integral for the effective 
action in Eq.~(\ref{SeffA}), we sample the field corresponding field configurations with a linear, 
nonlocal stochastic partial differential equation with additive noise. This generalizes 
to this spatially anisotropic nonequilibrium case, the commonly used stochastic equations in 
equilibrium Landau-Ginzburg theories~\cite{ma1, ma2,halperin,tauber}, allowing us to discuss 
the temporal behavior of the system; see, for example, Ref.~\cite{TT}. Specifically, 
we assume that $\xi_{i}(t,z,\x)$ is a genuine Gaussian-Markovian noise:
\begin{widetext} 
\begin{align}\label{eq:correlationxi}
\langle \xi_{i}^{(k)}(t,z,\x) \xi_{j}^{(k)}(t',z',{\x}')\rangle &=2\Upsilon \delta_{ij} \, \delta(t-t')
\, \delta(z-z') \, \delta^{d}(\x-{\x}'),
\end{align}
where $\langle\hdots\rangle$ means an average over all possible realizations of the noise. The corresponding 
stochastic equation sampling the field configurations $\varphi_{i}^{(k)}(t,z,\x)$ with weight  
$S_{\textrm{eff}}\left(\varphi_{i}^{(k)},j_{i}^{(k)}\right)$ is then given by the generalized
Langevin equation: 
\begin{align}\label{eq:langevintau}
\frac{\partial}{\partial t}\varphi_{i}^{(k)}(t,z,\x)=-\Upsilon\left.
\frac{\delta S_{\textrm{eff}}\left(\varphi_{i}^{(k)},j_{i}^{(k)}\right)}{\delta \varphi_{i}^{(k)}(z,\x)}
\right|_{\varphi_{i}^{(k)}(z,\x)=\varphi_{i}^{(k)}(t,z,\x)}+\xi_{i}^{(k)}(t,z,\x).
\end{align}
This equation is similar to the one that, after a coarse-grained procedure, describes the relaxational dynamics 
of classical nonequilibrium systems. In our case, $\Upsilon = 1$. Performing the functional derivatives, 
the generalized Langevin equation can be written as: 
\begin{align}\label{eq:langevintau44}
\frac{\partial}{\partial t}\varphi_{i}^{(k)}(t,z,\x)
+ \left(-\frac{\partial^{2}}{\partial z^{2}} - \Delta + m_{0}^{2}\right) \varphi_{i}^{(k)}(t,z,\x)
- \frac{\varrho^{2}}{L^{2}} \, \sum_{s=1}^{k}\int_{0}^{L}dv'\,\varphi_{s}^{(k)}(t,v',\x) 
= \xi_{i}^{(k)}(t,z,\x) + j_{i}^{(k)}(t,z,\x).
\end{align}
\end{widetext}

Once we discuss the two-point correlation function for large values of $L$, the limit of integration over $v'$ can be replaced from $[0,L]$ to $[0,z]$, assuming large values $z$. To deal with the nonlocal term, we employ a fractional derivative formalism, similar to the one 
used in studies of anomalous diffusion in transport processes through a disordered medium~\cite{topical}.
Specifically, we use the Riemann-Liouville fractional integrodifferential operator of order $\alpha$, 
$D_{a}^{\alpha}$~\cite{rr}. Let $f \in L^{1}[a,b]$ and $0<\alpha \leq1$; then $D_{a}^{\alpha}f$ exists 
almost everywhere in $[a,b]$, with $D_{a}^{\alpha}f$ defined by~\cite{rr}:
\begin{equation}
D_{a}^{\alpha}f(v)=\frac{1}{\Gamma(\alpha)}\int_{a}^{v} ds\,f(s)(v-s)^{\alpha-1} . 
\end{equation}
Therefore, the nonlocal term is given in terms of $D_{a}^{\alpha}$ as: 
\begin{align}\label{eq:langevintau444}
D^{1}_{0}\left(\varphi_{i}^{(k)}(t,z,\x) + \sum_{s=1,  s \neq i}^{k}\varphi_{s}^{(k)}(t,z,\x)\right).
\end{align}
The operator $D_{a}^{\alpha}f \equiv \frac{d^{\alpha}f(x)}{d|x|^{\alpha}}$  possesses a well-defined Fourier transform,
namely 
\begin{equation}
\mathcal{F}\left[\frac{d^{\alpha}f(x)}{d|x|^{\alpha}}\right] = - |k|^{\alpha}f(k), 
\hspace{0.5cm} \text{for} \hspace{0.5cm}
1\leq\mu<2.
\end{equation}
%
%

\noindent
We define the the Fourier transform on the time and spatial coordinates of a generic function 
$g(t,z,\x)$ by $\tilde{g}(\omega,q_{z},\q_perp) = \mathcal{F}_{t,z,\x}[g(t,z,\x)]$,
where $q_z = q_z(n) = n \pi/L,\; n\in \mathbb{Z}$. The Langevin equation in terms of the
Fourier-transformed functions is then given by:
\begin{align}
&\left[- i \omega + \biggl(\mathbf{q}_{\perp}^{2} + q_{z}^2 + m_{0}^2+ k\varrho^{2}   |q_{z}|\biggr)\right]
\tilde{\varphi}_{i}^{(k)}(\omega, q_{z}, 
\mathbf{q}_{\perp}) \nonumber\\ 
&\hspace{1.0cm} = \tilde{\xi}_{i}^{(k)}(\omega, q_{z}, \mathbf{q}_{\perp}) 
+ \tilde{j}_{i}(\omega, q_{z}, \mathbf{q}_{\perp}), 
\end{align}
in which we again assumed for each moment of the partition function equal field and source functions, $\phi_{i}^{(k)}(\x) = 
\phi_{j}^{(k)}(\x)$ and $j_{i}^{(k)}(\x)=j_{j}^{(k)}(\x)$. From this, one can compute the 
the dynamic susceptibility $\chi_{0}^{(k)}(\omega,,q_{z}, \mathbf{q}_{\perp})$, which is given 
by the response propagator $G_{0}^{(k)}(\omega,q_{z}, \mathbf{q}_{\perp})$:
\begin{equation}
\hspace{-0.25cm}G_{0}^{(k)}(\omega,q_{z}, \mathbf{q}_{\perp}) = 
\frac{1}{- i \omega + 
\left(\mathbf{q}_{\perp}^{2} + q_{z}^2 + m_{0}^2+ k\varrho^{2}  |q_{z}|\right)}.
\end{equation}

Near criticality in the pure-system, \textit{i.e.}, for $\varrho=0$, three critical exponents 
of the Gaussian model can be obtained: the two static exponents $\nu=\frac{1}{2}$ and $\eta=0$, and 
the dynamical exponent $z=2$. Using the principle of causality, contour integration leads to the following:for
the $G_{0}^{(k)}(t,q_{z}, \mathbf{q}_{\perp}) = \mathcal{F}^{-1}_{t} G_{0}^{(k)}(\omega,q_{z}, \mathbf{q}_{\perp})$:
\begin{align}
G_{0}^{(k)}(t,q_{z}, \mathbf{q}_{\perp}) &= \mathcal{F}^{-1}_{t} G_{0}^{(k)}(\omega,q_{z}, \mathbf{q}_{\perp})
\nonumber \\
&= \theta(t) \, e^{-\bigl(\mathbf{q}_{\perp}^{2} + q_{z}^2 + m_{0}^2+ k\varrho^{2}  |q_{z}|\bigr)t}, 
\end{align}
where $\theta(t)$ is the Heaviside theta function. Clearly, the function $G_{0}^{(k)}(t,q_{z}, \mathbf{q}_{\perp})$ 
decays exponentially to zero as $t \rightarrow \infty$. 

The next step is to find the Gaussian dynamic correlation function. Using the noise correlator 
in Fourier space for large $L$ we get
\begin{align}
\langle\tilde{\varphi}^{(k)}(&\omega, q_{z}, 
\mathbf{q}_{\perp})\tilde{\varphi}^{(k)}(\omega', q'_{z}, 
\mathbf{q'}_{\perp})\rangle=(2\pi)^{d+1}\delta(\omega+\omega')\nonumber\\
&\times
\delta(q_{z}+ q'_{z})
\delta(\mathbf{q}_{\perp}+\mathbf{q'}_{\perp})C^{(k)}_{0}(\omega, q_{z}, 
\mathbf{q}_{\perp}),
\end{align}
where
\begin{equation}
C_{0}^{(k)}(\omega, q_{z}, 
\mathbf{q}_{\perp})=2 \bigl(G^{(k)}_{0}(\omega,q_{z}, \mathbf{q}_{\perp})\bigr)^{2}
\end{equation}
is called the dynamical structure factor. The temporal correlation decays exponentially, 
with a modified relaxation rate due to the disorder. An experimental accessible quantity 
is the static structure factor $C_{0}^{(k)}( q_{z}, 
\mathbf{q}_{\perp})$, defined as 
\begin{equation}
C_{0}^{(k)}( q_{z}, 
\mathbf{q}_{\perp})=\frac{1}{2\pi}\int_{-\infty}^{\infty} d\omega\,C_{0}^{(k)}(\omega, q_{z}, 
\mathbf{q}_{\perp})  ,
\end{equation}
from which one can find the correlation lengths in the model. Since the disorder is anisotropic, 
the behavior of the system is different for distinct directions.  
In the Gaussian approximation in four-dimensional space, using a Fourier representation
for $G^{(k)}_{0}(z-z',\x-\y)$ one can show that
\begin{align}
&G_{0}^{(k)}(z-z',\x-\y)=\frac{1}{(2\pi)^{2}}\frac{1}{|\x-\y|}\int_{0}^{\infty} dq_z\nonumber\\&\times 
e^{
-|\x-\y|\sqrt{q^{2}_{z}+m_{0}^{2}+k\varrho^{2}q_{z}}} \; \cos\bigl(q_z(z-z')\bigr).
\end{align}
Defining the quantity $\varsigma = {\varrho^{2}}/{2\,m_{0}}$, we can write
\begin{align}
&G_{0}^{(k)}(z-z',\x-\y)=\frac{1}{(2\pi)^{2}}\frac{m_{0}}{|\x-\y|}
e^{-m_{0}|\x-\y|}
\nonumber\\&\times \int_{0}^{\infty} du \, e^{\sqrt{u^{2}+2k\varsigma u+1}} \, \cos  \left(m_{0}u(z-z')\right).
\end{align}
It is not possible to express this integral in terms of known functions, but we can circumvent 
this difficulty in the following way. We recall that the contribution of the terms of the 
series representation for the quenched free-energy given by
\begin{align}
\mathbb{E}\bigl[W(j,h)\bigr]=\sum_{k=1}^{\infty} c(k)\,\mathbb{E}\,[(Z(j,h))^{\,k}],
\end{align}
where $c(k)=\frac{(-1)^{k+1}}{k k!}$. For small $k$ such that $k\varsigma\rightarrow 0$, 
we can write, for large 
$(|\x-\y|^{2}+|z-z'|^{2})$ that the correlation function in a specific moment is given by
\begin{align}
G_{0}^{(k)}(z-z',\x-\y)=\frac{1}{\sqrt{8\pi^{5}}}\frac{\sqrt{m_{0}}\,e^{-m_{0}\sqrt{|z-z'|^{2}
+|\x-\y|^{2}}}}{\Bigl(|z-z'|^{2}+|\x-\y|^{2}\Bigr)^{\frac{3}{4}}}.
\end{align}
The contributions of these terms are the usual ones, for which the bulk correlation length 
$\xi=m_{0}^{-1}$ can be defined.  However, since $m_{0}^{2}>0$, there is no long-range order. 
Nevertheless, the existence of the long-range order can be obtained from the series representation 
of the quenched free-energy. 

For any real number $\kappa$, let $\lfloor\kappa\rfloor$ denote the largest 
integer $\leq \kappa$, that is, the integer $r$ for which $r\leq \kappa<r+1$.
We are interested in the critical moment of the partition function, which is the
$k_c=\lfloor\frac{2m_{0}}{\varrho^{2}}\rfloor$ moment. For this $k_{c}$-th moment,
the two-point correlation function has the form
\begin{align}
\label{grif}
G_{0}^{(k_{c})}(z-z',\x-\y)=\frac{1}{(2\pi)^{2}}\frac{e^{-m_{0}|\x-\y|}}{|z-z'|^{2}+|\x-\y|^{2}}.
\end{align}
This expression reflects the spatial anisotropy due to disorder. In the $k_c$-th moment, 
it is an explicit manifestation of generic scale invariance, as $G_{0}^{(k_{c})}(z-z',\x-\y)$
presents power-law decay in $z-z'$.


In the next section, we compute the mass in the one-loop approximation with anisotropic disorder in each moment of the partition function.

\section{Disorder effects in the one-loop correction to the mass}
\label{sec:thermal mass2}

The renormalized mass squared $m_{R}^{2}(L,\varrho,k)$ in the $k^{th}$ moment
is similar to the case discussed in Section~\ref{sec:thermal mass}, given by 
the contributions of a tadpole and a bubble diagram: 
\begin{align}
m_{R}^{2}(L,\varrho,k) &= m_{0}^{2} + \delta m^{2}_{0}
+ 3 \, \Delta m^{2}_{1}(L,\varrho,k)\nonumber\\
&+ 9 \, \Delta m^{2}_{2}(L,\varrho,k),
\end{align}
where $3$ and $9$ are symmetry factors, and again a mass counterterm $\delta m^{2}_{0}$
was introduced. Let us first discuss the contribution from tadpole diagram 
$\Delta m^{2}_{1}(L,\varrho,k)$ using the analytic regularization procedure discussed in 
Sec.~II. For $s\in \mathbb{C}$, $\Delta m^{2}_{1}(L,\varrho,k)$ can be obtained by the
analytic continuation $\Delta m^{2}_{1}(L,\varrho,k) = 
\Delta m^{2}_{1}(L,\varrho,k,\mu,s)|_{s=1}$,  with
\begin{align}
&\Delta m^{2}_{1}(L,\varrho,k,\mu,s) = \frac{\lambda(\mu,s)\,L}{2^{d+1}\pi^{\frac{d}{2}+1}\Gamma\bigl(\frac{d}{2}\bigr)}
\nonumber\\
&\times\int_{0}^{\infty}dp\,p^{d-1} \sum_{n\in \mathbb{Z}}
\left[\pi n^{2}+\frac{L}{2}k\varrho^{2}|n|+\frac{L^{2}}{4\pi}\Bigl(p^{2}+m_{0}^{2}\Bigr)
\right]^{-s}, 
\end{align}
where a trivial angular part of the integral was performed, and $\lambda(\mu,s)=\lambda_{0}(\mu^{2})^{s-1}$, 
where $\mu$ has dimension of mass. As in the case $\varrho = 0$ this function 
is defined in the region where the above integral converges, $\Re(s)>s_{0}$. 
Again, the contribution from the bubble diagram (self-energy) can be obtained from 
the tadpole:
\begin{equation}
\Delta m^{2}_{2}(L,\varrho,k)=\biggl[-\frac{\rho^{2}(\mu,s)}{\lambda(\mu,s)}
\, \Delta m^{2}_{1}(L,\varrho,k,\mu,s)\biggr]_{s=2},
\end{equation}
where $\rho(\mu,s)=\rho_{0}(\mu^{2})^{s-2}$.

After a Mellin transform and performing the $p$ integral, Eq.~(53) can be written 
as:
\begin{align}
\Delta m^{2}_{1}&(L,\varrho,k,\mu,s) = \frac{\lambda(\mu,s)}{4\pi\Gamma(s)}\biggl(\frac{1}{L}\biggr)^{d-1}
\int_{0}^{\infty}dt\,t^{s-\frac{d}{2}-1}\nonumber\\ &\times
\sum_{n\in \mathbb{Z}} e^{-\left (\pi\,n^{2}+
\frac{L}{2}k\varrho^{2}|n|+
{m_{0}^{2}L^{2}}{/4\pi}\right)t}.
\end{align}
One notices here that the anisotropic disorder introduced a contribution involving $|n|$ into 
the correlation function, which was not present in the one-loop correction for the pure-system discussed in
Sec.~\ref{sec:thermal mass}; see e.g. Eq.~(\ref{m1-mu-s}). This means that one needs to use 
a strategy different from that used in Sec.~\ref{sec:thermal mass} to deal with the 
renormalization of the one-loop correction in the present case. Specifically, we split the summation 
into $n=0$ and $n\neq 0$ contributions. We start with $\Delta m^{2}_{1}(L,\varrho,k,\mu,s)|_{n=0}$: 
\begin{align}
\Delta m^{2}_{1}(L,\varrho,k,\mu,s)|_{n=0}&=\frac{\lambda(\mu,s)}{4\pi\Gamma(s)}\biggl(\frac{1}{L}\biggr)^{d-1}A(s,d),
\end{align}
where 
\begin{align}
A(s,d) = \int_{0}^{\infty}dt\,t^{s-\frac{d}{2}-1} \, e^{- m_{0}^{2}L^{2}t/4\pi}.
\end{align}
For some $d$ and $s$, this integral has infrared divergence.
Different methods for infrared regularization have been discussed in the literature;
see, for example, Ref. \cite{lo1}. Here we implement another approach to deal with 
this infrared divergence~\cite{gelfand}. The integral $A(s,d)$ is defined for 
$\Re(s)>\frac{d}{2}$, and can be analytically continued to $\Re(s)>\frac{d}{2}-1$ 
for $s\neq \frac{d}{2}$. We write a regularized quantity $A_{R}(s,d)$ as
\begin{align}\label{eq:55}
&A_{R}(s,d)=\int_{0}^{1}dt\, t^{s-\frac{d}{2}-1}
\left( e^{- m_{0}^{2}L^{2}t/{4\pi}} - 1 \right)
\nonumber\\
&+ \int_{1}^{\infty}dt\, t^{s-\frac{d}{2}-1} \,
e^{- m_{0}^{2}L^{2}t/{4\pi}}
+ \frac{1}{\bigl(s-\frac{d}{2}\bigr)},
\end{align}
which is valid for $\Re(s)>\frac{d}{2}$. For $\Re(s)>\frac{d}{2}-1$ and $s\neq \frac{d}{2}$, the right-hand side exists and defines a regularization of the original integral.
Next, we consider~$\Delta m^{2}_{1}(L,\varrho,k,\mu,s)|_{n\neq 0}$:
\begin{align}\label{eq:56}
&\Delta m^{2}_{1}(L,\varrho,k,\mu,s)|_{n\neq 0}=\frac{\lambda(\mu,s)}{2\pi\Gamma(s)}\biggl(\frac{1}{L}\biggr)^{d-1}\int_{0}^{\infty}dt\,t^{s-\frac{d}{2}-1}\nonumber\\
&\times\sum_{n=1}^{\infty} e^{ - \pi \left(n^{2}+
{k L\varrho^{2}n}/{2\pi} + {m_{0}^{2}L^{2}}{/4\pi^{2}}
\right)t}.
\end{align}
We make use of the result obtained in the Appendix. In the series representation for the free-energy
with $k=1,2,..$, we have that for the moments of the partition function such that $k_{(q)}=\lfloor(\frac{2\pi q}{L})
\frac{2}{\varrho^{2}}\rfloor$, $(q \in \mathbb{N})$ are critical. This result is similar to the one
obtained in the Dicke model, where there is a quantum phase transition when the couplings between the raising 
and lowering off-diagonal operators and the bosonic mode, the energy gap between the energy eigenstates of the 
two-level atoms, and the frequency of the bosonic mode satisfy a specific constrain~\cite{dicke,aparicio1,aparicio2,aparicio3,ciuti}. 
Since we are interested in critical behavior, we will restrict our attention to the set of
critical moments. This means that one can replace the dependence $(\varrho,k)$ by $q$, so that:
\begin{align}
&\Delta m^{2}_{1}(L,q,\mu,s)|_{n\neq 0}=\frac{\lambda(\mu,s)}{2\pi\Gamma(s)}\biggl(\frac{1}{L}\biggr)^{d-1}
\int_{0}^{\infty}dt\,t^{s-\frac{d}{2}-1}\nonumber\\
&\times e^{-\pi \left( {m_{0}^{2}L^{2}}/{4\pi^{2}} - q^{2} \right)t }  
\sum_{n=1}^{\infty} e^{-\pi (n+q)^{2}t}.
\end{align}

Finally, let us show that $\Delta m^{2}_{1}(L,q,\mu,s)|_{n\neq 0}$ and also 
$\Delta m^{2}_{2}(L,q,\mu,s)|_{n\neq 0}$ are written in terms of the Hurwitz-zeta function. 
A simple calculation shows that choosing $q$ such that $q_{0}=\lfloor\frac{m_{0}L}{2\pi}\rfloor$, 
the quantity $\Delta m^{2}_{1}(L,q,\mu,s)|_{n\neq 0}$,  is given by
\begin{align}\Delta m^{2}_{1}(L,q_{0},\mu,s)|_{n\neq 0} &= \frac{\lambda(\mu,s)}{2\pi\Gamma(s)}
\biggl(\frac{1}{L}\biggr)^{d-1}
\int_{0}^{\infty}dt\,t^{s-\frac{d}{2}-1}\nonumber\\
&\times \sum_{n=1}^{\infty} e^{-\pi \left(n+q_{0}
\right)^{2}t}.
\end{align}
With the special choice $q_{0}=\lfloor\frac{m_{0}L}{2\pi}\rfloor$ we obtain the critical value of $k_c$, 
which was used to obtain Eq. (\ref{grif}). We interpret this result in the following way. 
In the infinite number of moments that defines the free-energy we obtain a subset of critical moments. 
In this subset, there is a particular set, for a specific value of $q$  that generates the tree level 
behavior. Going back to the above integral, this simplification allows one to write $\Delta m^{2}_{1}(L,q_{0},\mu,s)|_{n\neq 0}$ as 
\begin{align}
&\Delta m^{2}_{1}(L,q_{0},\mu,s)|_{n\neq 0}=\frac{\lambda(\mu,s)}{2\pi\Gamma(s)}\biggl(\frac{1}{L}\biggr)^{d-1}
\nonumber\\
&\times \Biggl[\int_{0}^{\infty}dt\,t^{s-\frac{d}{2}-1}
\sum_{n=0}^{\infty} e^{-\pi \bigl(n+q_{0}
\bigr)^{2} t} - A_{R}(s,d)\Biggr].
\end{align}

Let us analyze the quantity $F_{d}(L,,q_{0},\mu,s)$, defined by
\begin{align}
F_{d}(L,q_{0},\mu,s)&=\Delta m^{2}_{1}(L,q_{0},\mu,s)|_{n\neq 0}\nonumber\\
&+\frac{\lambda(\mu,s)}{2\pi\Gamma(s)}\biggl(\frac{1}{L}\biggr)^{d-1}A_{R}(s,d).
\end{align}
Using a inverse Mellin transform, we can write $F_{d}(L,q_{0},\mu,s)$ as:
\begin{align}
F_{d}(L,q_{0},\mu,s)&=\lambda(\mu,s)\biggl(\frac{1}{L}\biggr)^{d-1}
\frac{\Gamma(s-\frac{d}{2})\pi^{\frac{d}{2}-s-1}}{2\Gamma(s)}\nonumber\\
&\times \zeta\bigl(2s-d,q_{0}\bigr), 
\end{align}
where the Hurwitz-zeta function $\zeta(z,a)$ is defined by 
\begin{equation}
\zeta(z,a) = \sum_{n=0}^{\infty}\frac{1}{\bigl(n+a\bigr)^{z}}, 
\,\,\,\,\,\,\,a\neq 0,-1,-2,...
\end{equation}
The series converges absolutely for $\Re(z)>1$. It is possible to find the analytic continuation, 
with a simple pole at $z=1$. For $d=3$, the contribution from the tadpole is finite, 
but the contribution from the self-energy is divergent. An important formula that must be used 
in the renormalization procedure is
\begin{equation}
\lim_{z\rightarrow 1}\biggl[\zeta(z,a)-\frac{1}{z-1}\biggr]=-\psi(a),
\label{psi}
\end{equation}
where $\psi(a)$ is the digamma function defined as $\psi(z)=\frac{d}{dz}\bigl[\ln\Gamma(z)\bigr]$.
Using the Hurwitz-zeta function and the integral $A_{R}(s,d)$, we can write:
\begin{align}
&\Delta m^{2}_{1}(L,q_{0},\mu,s)|_{n\neq 0}=\frac{\lambda(\mu,s)}{2\Gamma(s)}\biggl(\frac{1}{L}\biggr)^{d-1}
\nonumber\\
&\times\Biggl[\pi^{\frac{d}{2}-s-1}
\Gamma\biggl(s-\frac{d}{2}\biggr)\zeta(2s-d,q_{0})-\frac{1}{\pi}A_{R}(s,d)\Biggr].
\end{align}
 
Next, we prove that for a fixed value of $q_{0}$ the renormalized squared mass vanishes 
for a family of $L's$. In low-temperature field theory we get the same result, 
\textit{i.e}, there are critical temperatures where the renormalized squared mass 
vanishes, namely:
\begin{align}
m_{R}^{2}(L,q_{0}) & = m_{0}^{2} + \delta m_{0}^{2} + 3\, \Delta m^{2}_{1}(L,1)|_{n=0}
\nonumber\\
&+9 \, \Delta m^{2}_{2}(L,2)|_{n=0} + 3\, \Delta m^{2}_{1}(L,q,1)|_{n\neq 0} 
\nonumber\\
& + 9 \, \Delta m^{2}_{2}(L,q,2)|_{n\neq 0} .
\end{align}
Defining the dimensionless quantities $b = m_0 L$, $\lambda_{1}=3\lambda_{0}$, and  
$\rho_{2}=\rho_{0}\sqrt{9}$, we can write the latter equation as:
\begin{align}\label{eq:66}
\frac{b^{d-1}}{m_0^{d-3}} &- \frac{\lambda_1}{4\pi}A_{R}(1,d) + \frac{\rho_2^2}{4\pi\mu^2}A_{R}(2,d)+\delta m_{0}^{2} 
\nonumber\\ 
& + \frac{\lambda_1\pi^{\frac{d}{2}-2}}{2} \Gamma \left(1-\frac{d}{2}\right) 
\zeta\left(2-d,\frac{b}{2\pi}\right)\nonumber\\ 
& - \frac{\rho_2^2\pi^{\frac{d}{2}-3}}{2\mu^2} \Gamma 
\left(2-\frac{d}{2}\right)\zeta\left(4-d,\frac{b}{2\pi}\right)=0.
\end{align}
Let us discuss the case $d=3$, in which case Eq. (\ref{eq:66}) becomes:

\begin{align}\label{eq:67}
&b^{2} - \frac{\lambda_1}{4\pi}A_{R}(1,3) + \frac{\rho_2^2}{4\pi\mu^2}A_{R}(2,3) - \lambda_1
\zeta\left(-1,\frac{b}{2\pi}\right)\nonumber\\  -&
\frac{\rho_2^2}{2\pi\mu^2}\lim_{d \to 3}\zeta\left(4-d,\frac{b}{2\pi}\right) + \delta m_{0}^{2} =0.
\end{align}
The contribution coming from $A_R(s,d)$ is irrelevant for large $m_{0}L$, as one can verify 
in Eq.~(\ref{eq:55}). Using the identity $(n+1) \zeta(-n,a) = - B_{n+1}(a)$,
where the $B_{n+1}(a)$ are the Bernoulli polynomials, we rewrite the Hurwitz-zeta function as

\begin{align}
\zeta\left(-1,\frac{b}{2\pi}\right) &=
- \left(\frac{b^2}{8\pi^2} - \frac{b}{4\pi} + \frac{1}{12}\right).
\end{align}
Using Eq. (\ref{psi}) we fix the counterterm contribution in the renormalization procedure. 
Then we have that Eq.~(\ref{eq:67}) becomes:
\begin{align}\label{eq:69}
&b^{2} + \lambda_1\left(\frac{b^2}{8\pi^2} - \frac{b}{4\pi} + \frac{1}{12}\right) 
+ \frac{\rho_2^2}{2\pi\mu^2}\psi\left(\frac{b}{2\pi}\right)\,=0.
\end{align}
Since $q_0=\lfloor\frac{b}{2\pi}\rfloor$, we can write
the digamma function as
\begin{align}
\psi(q_0 + \sigma) = \psi(\sigma) + \sum_{q=1}^{q_0}\frac{1}{\sigma + q}\,,
\end{align}
where $\sigma$ is the noninteger part of $\frac{b}{2\pi}$. With $\sigma < 1$ we can 
use a Taylor's series and write Eq. (\ref{eq:69}) as: 
\begin{align}\label{eq:b}
& b^{2} + \lambda_1\left(\frac{b^2}{8\pi^2} - \frac{b}{4\pi} + \frac{1}{12}\right)\nonumber \\ 
& + \frac{\rho_2^2}{2\pi\mu^2}\left(-\frac{1}{\sigma} 
- \gamma + \frac{\pi^2}{6}\sigma + H_{q_0}^{(1)} + \sigma H_{q_0}^{(2)}\right)=0,
\end{align}
where $H_{q_0}^{(1)}$ and $H_{q_0}^{(2)}$ are the generalized harmonic numbers. The above equation 
has zeros for different values of $L$.  
\begin{figure}[ht!]\label{fig:b}
\includegraphics[scale=0.67]{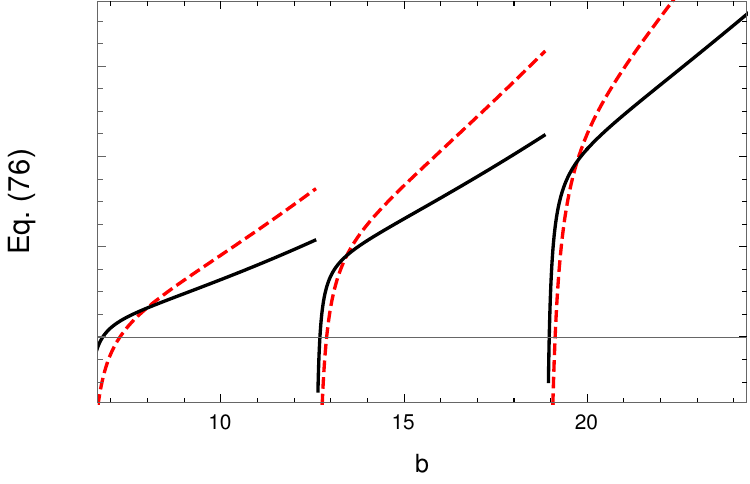}
\label{fig:2}
\caption{Plot of Eq. (\ref{eq:b}) as a function of $b = m_0L$ for two different values of $\lambda_0$ (once $\rho_0^2 = 2m_0^2\lambda_0$): $\lambda_0 = 1$ (continuous black) and $\lambda_0 = 3$ (dashed red). We set $\mu^2=m_0^2$.}
\end{figure}

In one-loop approximation we proved that in the set of moments that defines the quenched free-energy there 
is a denumerable collection of moments that can develop critical behavior. With the bulk in the ordered phase, 
in these moments temperature or finite size effects lead the moments from the ordered to a disordered phase.
Also, in the set of moments, there appears a large number of critical temperatures.

\section{Conclusions}\label{sec:conclusions}

In this paper, we studied quantum and disorder fluctuations in systems with quenched disorder. We employed a Euclidean quantum $\lambda\varphi^{4}_{d+1}$ model with the disorder linearly 
coupled to the scalar field $\varphi$. We used the equivalence between the model defined in a $d$-dimensional 
space with imaginary-time with the classical model defined on a space ${\mathbb R}^{d}\times S^{1}$. 
The physical quantity of interest, the disorder-averaged free-energy, was computed as a series of the 
moments of the partition function. Each moment is characterized by an effective action of a multiplet of fields. 
The effective actions contain nonlocal terms generated by the anisotropic disorder. We computed the tree level
two-point correlation functions for the fields of a given moment employing stochastic differential equations 
with additive white noise, with each equation driven by the effective action of the corresponding moment. We employed 
the method of fractional derivatives to deal with the nonlocal terms that appear in the stochastic 
equations. The two-point correlation functions for the critical moment display spatial anisotropy 
due to disorder, with manifest generic scale invariance. Finally, we proved at the one-loop order that,
with the bulk in the ordered phase, there is a denumerable set of moments that lead the system to the
critical regime. In these critical moments there appears a large number of critical compactified 
lengths (or temperatures).


In the study of complex spatial patterns and structures in nature, there appears the idea of self-organized 
criticality~\cite{bak1,bak2}. The authors of these references suggest that fractal structures and $1/f$ noise 
are common characteristics of irreversible dynamics of a critical state, without a fine-tuning of external 
parameters. The algebraic decay of the correlation function in space and time for generic parameters is 
called generic scale invariance. Our Eqs.~(\ref{grif}) and (\ref{eq:b}), and Fig.~2 are a manifestation of 
generic scale invariance in an equilibrium system.

A natural continuation of this work is to discuss the random-mass model. The main difference between 
the random field and the random-mass models is that after the coarse-graining, in the former the 
disorder modifies the Gaussian contribution of the model, whereas in the latter the disorder affects 
the non-Gaussian contribution to the effective action. Another possibility is to discuss a model 
with a continuous symmetry, where the spontaneous symmetry breaking leads to Goldstone bosons, 
softy modes that naturally manifest generic scale invariance.
Both subjects are under investigation by the authors.

\begin{acknowledgments} 
This work was partially supported by Conselho Nacional de Desenvolvimento Cient\'{\i}fico e Tecnol\'{o}gico (CNPq), 
grants nos. 305894/2009-9 (G.K.) and 303436/2015-8 (N.F.S.), INCT F\'{\i}sica Nuclear e Apli\-ca\-\c{c}\~oes, grant no.
464898/2014-5  (G.K), and Funda\c{c}\~{a}o de Amparo \`{a} Pesquisa do Estado de S\~{a}o Paulo (FAPESP), grant no. 
2013/01907-0 (G.K). G.O.H thanks Coordena\c{c}\~ao de Aperfei\c{c}oamento de Pessoal de Nivel Superior (CAPES) for
a Ph.D. scholarship.  
\end{acknowledgments}
\appendix

\section{The critical moments of the partition function}\label{ap:a10}

In this appendix we discuss the contribution coming from  particular moments of the partition function. 
We will prove that these moments lead to critical domains in which the mass $m_{R}^{2}(L,\varrho,k)$ vanishes. 
 Starting from Eq.~(53) and splitting the contributions of $n=0$ and $n\neq0$, for operational purposes, we get:
\begin{align}
&m_{R}^{2}(L,\varrho,k) = m_{0}^{2} + \delta m_0^2 + 6\Delta m^{2}_{1}(L,\varrho,k,1)|_{n=0}\nonumber\\
&+ 18 \Delta m^{2}_{2}(L,\varrho,k,2)|_{n=0} + 6 \Delta m^{2}_{1}(L,\varrho,k,1)|_{n\neq 0}\nonumber\\ 
&+ 18 \Delta m^{2}_{2}(L,\varrho,k,2)|_{n\neq 0},
\end{align}
where $\Delta m ^2_1$ is the tadpole contribution and $\Delta m ^2_2$ is the self-energy contribution, with 
the parameters $\mu$ and $s$ being related to the regularization procedure. $\Delta m ^2_1 (L,\varrho, k, \mu,s)$ 
is given by:
\begin{align}
&\Delta m^{2}_{1}(L,\varrho,k,\mu,s)=\Delta m^{2}_{1}(L,\varrho,k,\mu,s)|_{n=0} \nonumber\\
&+\Delta m^{2}_{1}(L,\varrho,k,\mu,s)|_{n\neq 0}.
\end{align}
Let us start with the $\Delta m^{2}_{1}(L,\varrho,k,\mu,s)|_{n\neq 0}$ contribution, which can be written as:
\begin{align}
\Delta m^{2}_{1}(L,\varrho,k,\mu,s)|_{n\neq 0} &=
\frac{\lambda(\mu,s)}{2\pi\Gamma(s)}\biggl(\frac{1}{L}\biggr)^{d-1} \nn \\
&\hspace{-0.5cm}\times \int_{0}^{\infty}dt\,t^{s-\frac{d}{2}-1} \, e^{- t \left(m^2_0 
-  k^2\varrho^{4}/4\right) L^2/4\pi } \nn\\
&\hspace{-0.5cm} \times \sum_{n=1}^{\infty} e^{- \pi t \left(n + L k\varrho^{2}/{4\pi}
\right)^{2} }.
\end{align}
This can be split into three contributions: 
\begin{align}
\Delta m^{2}_{1}(L,\varrho,k,\mu,s)|_{n\neq 0} &= -\frac{\lambda(\mu,s)}{2\pi\Gamma(s)}\
\left(\frac{1}{L}\right)^{d-1} \nn \\
& \times  \sum^{3}_{i=1} I_{i}(L,\varrho,k,\mu,s).
\end{align}
where 
\begin{align}
I_{1}(L,\varrho,k,\mu,s) &= \int_{0}^{\infty}dt\,t^{s-\frac{d}{2}-1}
e^{ - t \, \left( m_{0}^{2}L^{2}/{4\pi}\right) }, \\
I_{2}(L,\varrho,k,\mu,s) &= \int_{0}^{\infty}dt\,t^{s-\frac{d}{2}-1}
\, e^{- t \, \left({m_{0}^{2}} - k^2\varrho^{2}/4\right)L^2/4\pi} \nonumber\\
&\times \sum_{n=1}^{\infty} e^{-\pi t\left(n - {L k\varrho^{2}}/{4\pi}\right)^2}, \\
I_{3}(L,\varrho,k,\mu,s)
&=\int_{0}^{\infty}dt\,t^{s-\frac{d}{2}-1}\Theta\biggl(t,\frac{L k\varrho^{2}}{4\pi}\biggr)
\nonumber\\
&\times e^{-t \left( m^2_{0} - k^2 \varrho^{4}/4\right)L^2/4\pi}.
\end{align}

Note that we used the theta series $\Theta(t;\alpha)$, in the above integral. Recall that for an arbitrary complex number $\alpha$ and also $t\in \mathbb{ C}$ with $\Re (t)>0$ 
the theta series $\Theta(t;\alpha)$ is defined as
\begin{equation}
\Theta(t;\alpha)=\sum_{n=-\infty}^{\infty} 
e^{-\pi t (n+\alpha)^{2}}.
\end{equation}
It is clear that $\Theta(t;\alpha)=\Theta(t;\alpha+1)$. One also has the identity:
\begin{align}
\Theta\Bigl(\frac{1}{t};\alpha\Bigl) &= \sqrt{t}\sum_{n=-\infty}^{\infty}
e^{ - \pi n^{2}t + 2\pi i n \alpha} 
\nonumber\\ 
&= \sqrt{t} \, e^{-\pi\alpha^{2}/t } \, \Theta\left(t;-{i\alpha}/{t}\right).
\end{align}\\
We will show that one has to select a specific set $k$'s in the summation of the series 
for the quenched free-energy. Let us split the integral into two regions. Since the theta series 
$\Theta(t;\alpha)$ is holomorphic in the half-plane $\Re(t)>0$, the $I_{3}(L,\varrho,k,\mu,s)$ 
contribution must be written as 
\begin{align}
I_{3}(L,\varrho,k,\mu,s)&=I_{3}^{(1)}(L,\varrho,k,\mu,s) +
I_{3}^{(2)}(L,\varrho,k,\mu,s),
\end{align}
where $I_{3}^{(1)}(L,\varrho,k,\mu,s)$ is given by
\begin{align}
I_{3}^{(1)}(L,\varrho,k,\mu,s)
&= \int_{1}^{\infty}dt\,t^{\frac{d}{2}-s-\frac{1}{2}} 
e^{- \left(m^2_{0} - k^2 \varrho^{4/4}\right)L^2/(4\pi t)} 
\nn \\
&\times \sum_{n=-\infty}^{\infty}
e^{ -\pi n^{2}t + i k L\varrho^{2}n/2}, \\
I_{3}^{(2)}(L,\varrho,k,\mu,s) &= \int_{1}^{\infty}dt\,t^{s-\frac{d}{2}-1}\Theta\biggl(t;\frac{L k\varrho^{2}}{4\pi}\biggr)\nonumber\\
&\times e^{-t ( m_{0}^{2} - k^2 \varrho^{4}/4)^2 L^2/4 \pi } .
\end{align}
The integral  $I_{3}^{(2)}(L,\varrho,k,\mu,s)$ converges absolutely for any $s$ and converges uniformly with respect to $s$
in any bounded part of the plane. Hence the integral represents an everywhere regular function of $s$. Concerning integral 
$I_{3}^{(1)}(L,\varrho,k,\mu,s)$, to guarantee the convergence we must choose 
$k_{(q)}=\lfloor(\frac{2\pi q}{L})\frac{2}{\varrho^{2}}\rfloor$ where $q$ is a natural number. Therefore, in the series 
representation for the free-energy with $k=1,2,..$ we have that for the moments of the partition function such that 
$k_{(q)}=\lfloor(\frac{2\pi q}{L})\frac{2}{\varrho^{2}}\rfloor$, 
where $(\frac{2\pi q}{L})$ are the positive Matsubara frequencies $\omega_{q}$, the system is critical. 
This is an interesting result, there is a critical set of moments in the series representation for the free-energy, 
after averaging over the quenched disorder. A more general proof using generalized Hurwitz-zeta functions is based on 
the fact that zeta function regularization with a meromorphic extension to the whole complex plane needs an eligible 
sequence of numbers~\cite{voros}.

\providecommand{\noopsort}[1]{}\providecommand{\singleletter}[1]{#1}%

\end{document}